\documentclass[12pt,a4paper]{conference}

\usepackage{fancyhdr}
\usepackage{graphicx,amsmath,amssymb,cite}
\usepackage{multind}
\makeindex{author} \makeindex{subject}

\newcommand{\beq}{\begin{equation}}
\newcommand{\eeq}[1]{\label{#1}\end{equation}}
\newcommand{\eeqn}{\end{equation}}

\newcommand{\beqa}{\begin{eqnarray}}
\newcommand{\eeqa}[1]{\label{#1}\end{eqnarray}}
\newcommand{\eeqan}{\end{eqnarray}}

\let\bar=\overbar

\newcommand{\Dslash}{\not{\hbox{\kern-4pt $D$}}}
\newcommand{\dslash}{\not{\hbox{\kern-2pt $\del$}}}

\newcommand{\msb}{{\bar{\ssstyle M \kern -1pt S}}}

\begin{document}

\Chapter{Quark model: recent issues}
           {Quark model: recent issues}{Fl. Stancu}
\vspace{-5 cm}\includegraphics[width=6 cm]{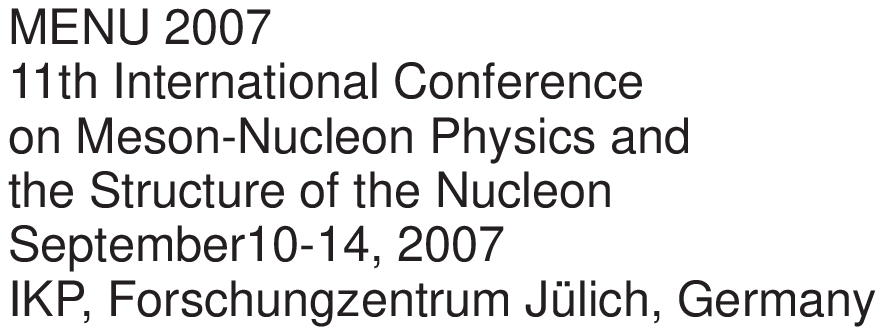}
\vspace{4 cm}

\addcontentsline{toc}{chapter}{{\it N. Author}} \label{authorStart}

\begin{raggedright}

{\it Fl. Stancu 
}\index{author}{Fl. Stancu}\\
Institute of Physics, B.5
\\
University of Li\`ege\\
Sart Tilman, B-4000 Li\`ege 1 \\
Belgium
\bigskip\bigskip

\end{raggedright}

\begin{center}
\textbf{Abstract}
\end{center}

In the first part I briefly survey 
recent issues in constituent quark models raised by the observation 
of unusual hadronic states.  In particular I  discuss 
the role of higher Fock components in the wave function of baryons and the  
possible interpretation of open charm and of new charmonium-type 
resonances as tetraquarks. In the second part I show support for the
quark model dynamics obtained in a model independent 
way from the $1/N_c$ expansion approach of QCD which proved to be successful in 
describing baryon properties.

\section{Introduction}

The organizers asked me to talk about recent issues in the quark model
(QM). This is a vast subject and I had to make a selective choice.
My talk contains two distinctive parts. The first is devoted to
specific issues in the QM related to  the recent observation  (since 2003)
of unusual hadronic states. In this context I 
present a few aspects of  the QM developments
in the light of the newly found  resonances.
The second part is devoted to a comparison between QM results and
the $1/N_c$ expansion approach of QCD, both being successful 
in baryon spectroscopy, the latter being closer to QCD and model
independent.

\section{The QM and the newly found resonances}

Here I refer to constituent quark (or potential) models. The basic assumptions 
are that the Hamiltonian consists
of a kinetic (non-relativistic or relativistic) part, a 
confinement part and a hyperfine 
interaction of a one-gluon exchange (OGE) type, a Goldstone boson exchange (GBE)
type or resulting from an instanton induced interaction (III), or 
a mixture of them. 

The quark models are generally successful in reproducing baryon spectra.
Relativistic effects turn out to be specially important in describing 
electromagnetic or axial form factors of light baryons. The strong 
decays of baryons
remain problematic in all potential models. The decay widths are generally 
underestimated in OGE or GBE models \cite{SMP} 
as well as in III models  \cite{METSCH}. 
These are results based on the description of the baryons as a system 
of three valence quarks. Till recently most of mesons were well described as
$q \bar q$ systems. 

The discovery of new exotic
resonances starting from 2003 brought new aspects into the standard
treatment of hadrons. These are: 1) higher Fock components in the 
wave function of some baryons 2) additional spin-orbit term in 
mesons with non-identical quark masses or the interpretation
of some of them as tetraquarks, for example. 
 
\subsection{Higher Fock components in baryon states}

The debate on the existence of pentaquarks lead to the study 
of the role of five quark components $(q^4 \bar q)$ in the 
wave function of the nucleon whenever there is a problem in
the description of a baryon as a $q^3$ system.
For example, the implication 
of such components has been analyzed in connection with 
experiments on parity violation in electron-proton scattering 
which suggest that the strangeness magnetic moment  $\mu_s$ of the
proton is positive. So far calculations gave either positive or
negative values. A positive value was obtained \cite{ZR}
by including in the wave function a positive parity component 
$ u u d s \bar s$ with one quark  in the 
$ u u d s $ subsystem  excited to the $p$-shell.    
The most favorable configuration  
is $[31]_O [22]_F[22]_S$, the same as for positive parity
pentaquarks \cite{FS} with a flavor-spin dependent GBE interaction.

Also, it is known that the QM, irrespective of the hyperfine interaction
included  in the Hamiltonian model, cannot explain the low 
mass of the $\Lambda$(1405) resonance. Recently dynamical calculations
based on the $q^4 \bar q$ configurations have been performed
\cite{TS}. These studies require an embedded $q^3$ pole in the
continuum and a coupling between $q^3$ and  $q^4 \bar q$ configurations
which remains an open problem.

Finally, there is the suggestion that a $uuds \bar s$ component
in the wave function of the N*(1535) resonance could lead to 
a larger coupling to N$\eta$ and N$\eta'$ channels \cite{ZOU},
in agreement to experiment.


\subsection{Tetraquarks}
Since 2003 an important number of exotic meson-like resonances have been
discovered. These are open charm resonances:  
$D_s(2317)$, $D_s(2460)$, $D_s(2690)$, $D_s(2860)$
and charmonium type (hidden charm) resonances:
$X(3872)$, $X(3940)$, $Y(3940)$, $Z(3930)$, 
$Y(4260)$, $Z^{\pm}(4433)$.

While for open charm resonances a canonical interpretation 
as $c \bar s$  systems is still possible
\cite{CJ,CTLS} through the addition of a spin-orbit
term which vanishes for equal quark and antiquark masses,
in the case of hidden charm resonances one has to assume
more complicated structures as: tetraquarks $(c \bar c)(q \bar q)$, 
$D \bar D^*$ molecules, hybrids, glueballs, etc. (for a review see 
for example Refs. \cite{ROSNER,SWANSON}). 
The tetraquark interpretation of X(3872) is quite attractive
\cite{MAIANI,HRS,VARENNA}. 
The quark model \cite{VARENNA} gives twice more states than the 
diquark model \cite{MAIANI}. However much work is still needed
in the framework of QM or other approaches in order to
understand the exotic hidden charm resonances.


\section{Compatibility of the quark model and the $1/N_c$ expansion approach}

The QM still remains the basic tool in hadron spectroscopy. In addition,
it has played 
an important role in the evolution of ideas towards QCD. However there
is no known way to derive the QM from QCD. Each constituent quark model
is based on a given Hamiltonian which contains a number of dynamical
assumptions. The results are obviously model dependent. Therefore it 
is very important to establish a connection between QM results and 
another approach, also successful in baryon spectroscopy, but model
independent and much more closely related to QCD. This is the 
$1/N_c$ expansion method described below. 


\subsection{The $1/N_c$ expansion method}

In 1974 't Hooft \cite{THOOFT}
extended QCD from SU(3) to SU($N_c$), where
$N_c$ is an arbitrary number of colors and suggested a perturbative
expansion in the parameter $1/N_c$, applicable to all QCD regimes. 
Witten applied 
the approach to baryons \cite{WITTEN} and derived power counting rules
which lead to a powerful $1/N_c$ expansion method to study static
properties
of baryons, as for example,  masses,  magnetic moments, 
axial currents, etc. The method is systematic and predictive.
It is based on the discovery that, in the limit
$N_c \rightarrow \infty$, QCD possesses an exact contracted
SU(2$N_f$) symmetry  \cite{GS,DM} where $N_f$ is the number
of flavors. This symmetry is
only approximate for finite $N_c$ so that corrections have to be added
in powers of $1/N_c$.

\begin{figure}[hb]
\begin{center}
\includegraphics[height=6 cm, width=8.5 cm]{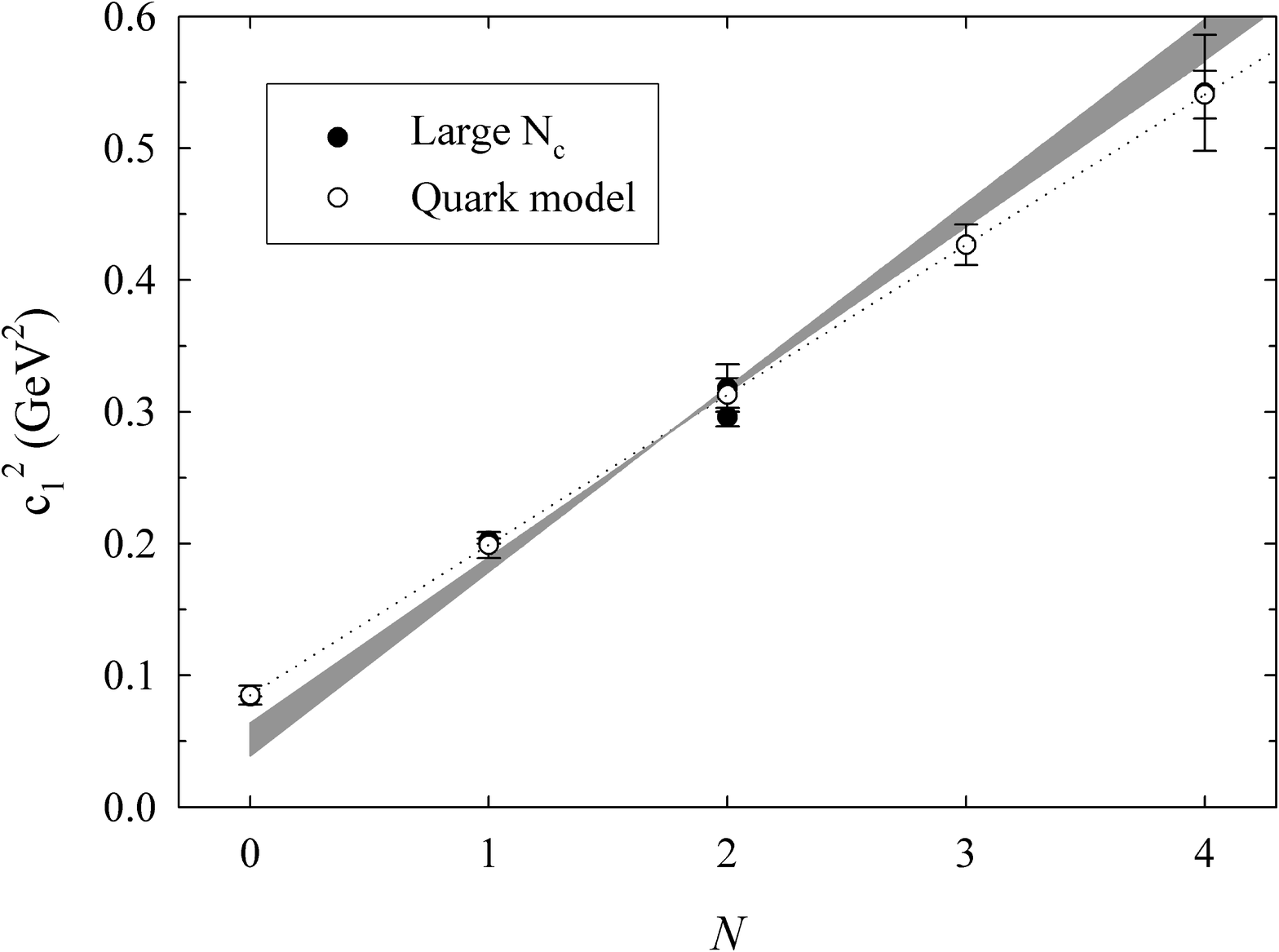}
\caption{$c^2_1$ from the $1/N_c$ expansion mass formula, 
Eq. (\ref{massoperator}), and the 
QM result, Eq. (\ref{totalmass}).
(for details on the large $N_c$ points see Ref. \cite{SBMS}).} 
\label{fig:coef1}
\end{center}
\end{figure}

In the $1/N_c$ expansion approach the mass operator 
has the general form
\begin{equation}
\label{massoperator}
M = \sum_{i} c_i O_i + \sum_{i} d_i B_i,
\end{equation}
where each sum extends over a finite number of terms.
The operators $O_i$ are invariants under SU(6) transformations
and the operators $B_i$ explicitly break SU(3)-flavor symmetry.
The coefficients $c_i$ and $d_i$ encode the quark dynamics
and are fitted
to the experimental data. In the case
of nonstrange baryons, only the operators $O_i$ contribute while
$B_i$ are defined such as their expectation values are zero.
The building blocks of $O_i$ and $B_i$ are the SU(6) generators: $S_i$
($i$ = 1,2,3) acting on spin and forming an su(2) subalgebra,
$T^a$ ($a$ = 1,...,8)
acting on flavor and forming an su(3) subalgebra, and $G^{ia}$ acting
both on spin and flavor subspaces. For orbitally excited states, also
the components $\ell_i$ of the angular momentum,
as generators of SO(3), and the tensor operator $\ell^{ij}$
are necessary to build $O_i$ and $B_i$.
Examples of $O_i$ and $B_i$ can be found in
Refs.~\cite{GSS,MS1,MS2,MS3}.
Each operator  $O_i$ or $B_i$ carries an explicit factor of
$1/N^{n-1}_c$
resulting from the power counting rules \cite{WITTEN}, where $n - 1$
represents the minimum of gluon exchanges to generate the operator.
In the matrix elements there are also compensating factors of $N_c$
when one sums coherently over $N_c$ quark lines. In practice
it is customary to drop higher order corrections of order $1/N^2_c$.

The discussion below concerns 
the coefficients $c_1$, $c_2$ and $c_4$ in Eq. (\ref{massoperator})
related to  the following operators
\begin{equation}
O_1 = N_c, ~~ O_2 = \ell_i S_i, ~~ O_4 = \frac{1}{N_c} S_i S_i.
\end{equation}
These are the spin-isospin independent,
the spin-orbit and the spin-spin operators respectively.
The analysis  is straightforward for resonances described by
symmetric states $[56, \ell]$.  
For mixed symmetric states $[70,\ell]$ the procedure
is more complicated due to the separation of the system
into a symmetric core and an excited quark.
In principle one can use a simpler approach in order to
avoid such separation \cite{MS4}.

\begin{figure}[hb]
\begin{center}
\includegraphics[height=6 cm, width=8.5 cm]{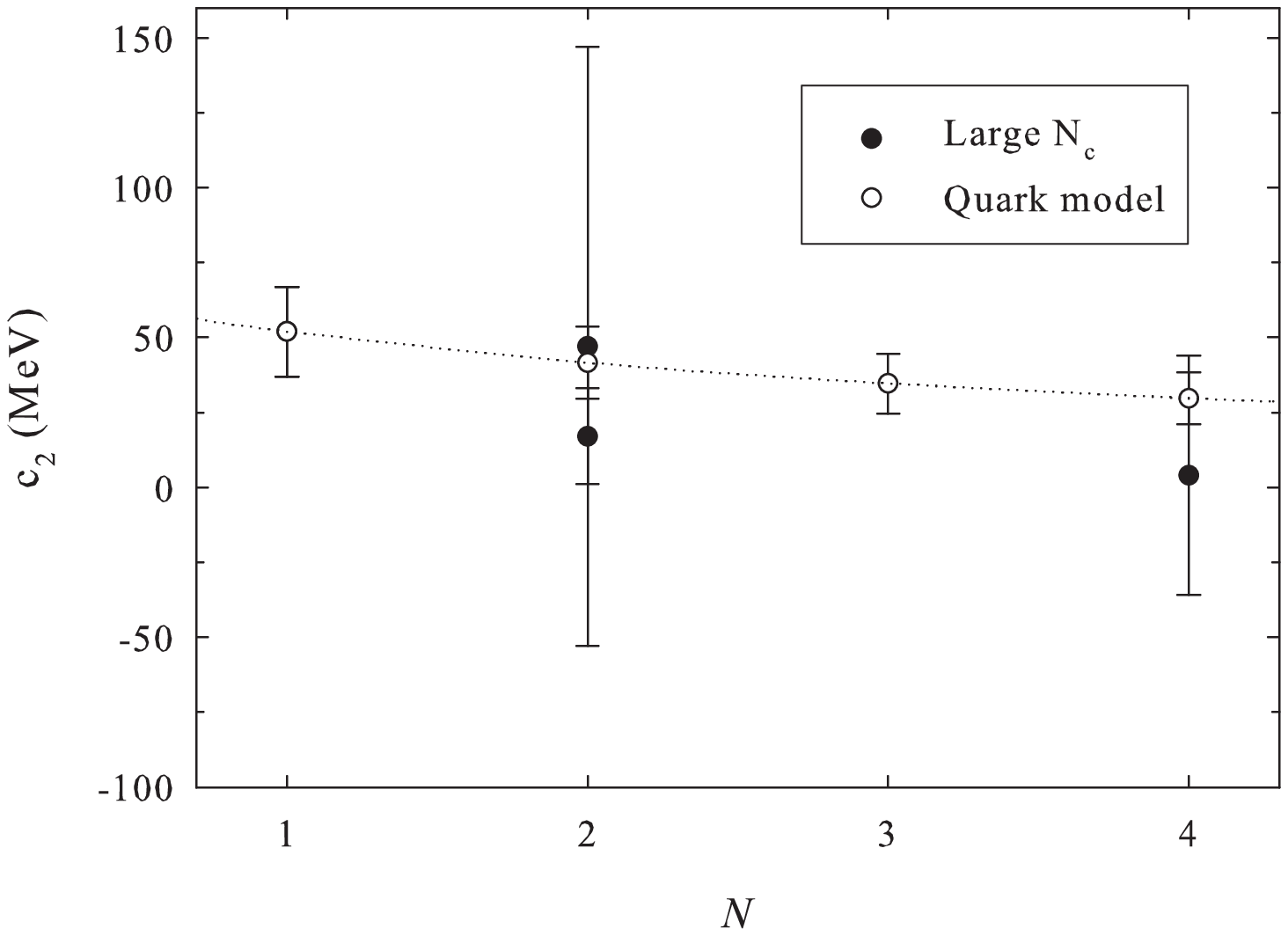}
\hspace{4 cm}
\includegraphics[height=6 cm, width=8.5 cm]{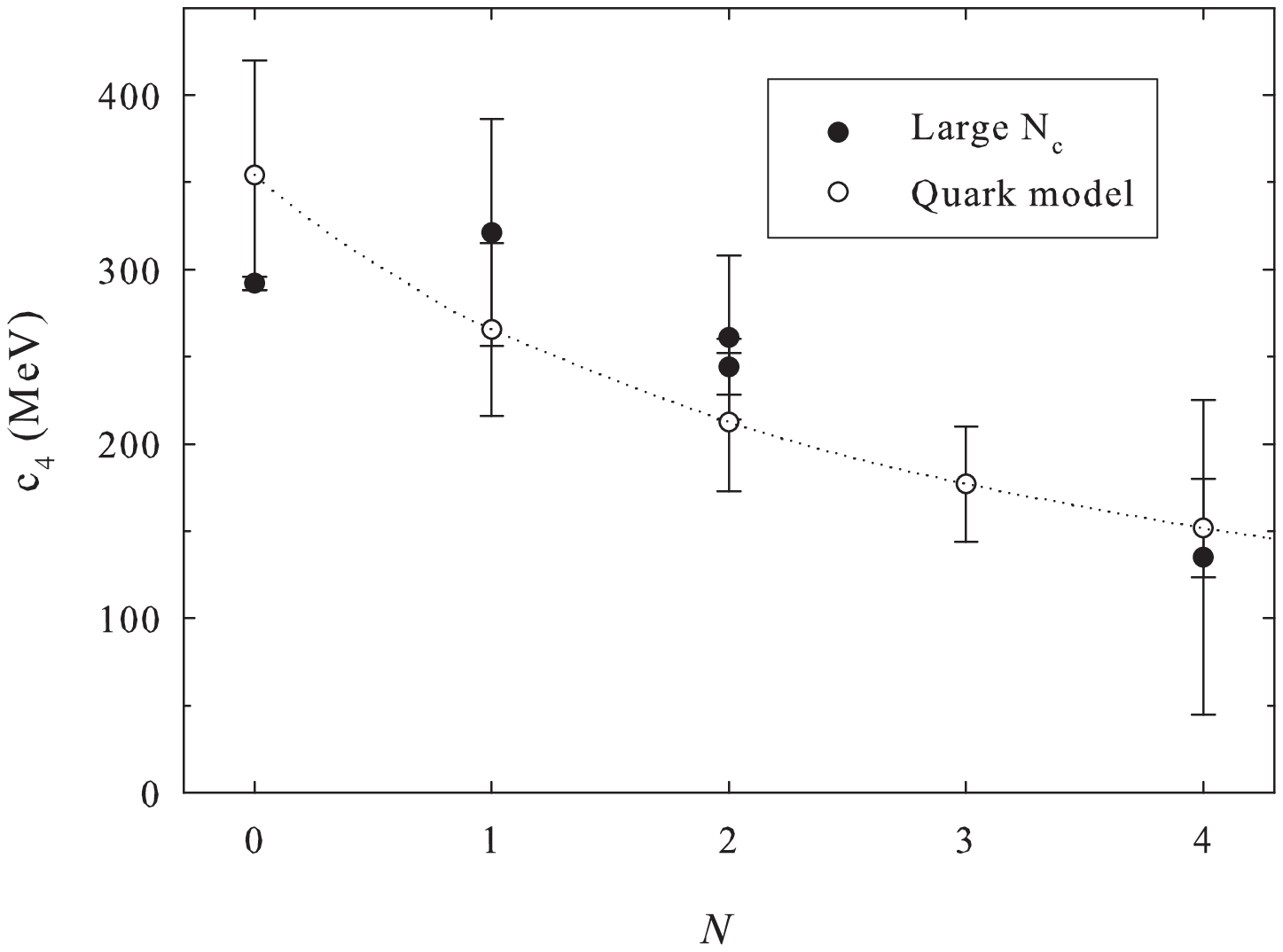}
\caption{The coefficients $c_2$ and $c_4$ in the large $N_c$ and 
QM approaches
(for details see Ref. \cite{SBMS}). The dotted line passes through
QM results.} \label{fig:coef2and4}
\end{center}
\end{figure}

For strange baryons one has
to include  both $O_i$ and $B_i$ operators in Eq. 
(\ref{massoperator}). The contribution to each strange quark to 
the mass, denoted by $\Delta M_s$, is given by 
\begin{equation}\label{break}
n_s ~\Delta M_s = \sum_{i} d_i B_i
\end{equation}
where $n_s $ is the number of strange quarks in a baryon.

\subsection{The quark model}

We follow the approach of Ref. \cite{SBMS} and 
start from the spinless Salpeter Hamiltonian
\begin{equation}\label{EXACT}
H = \sum_{i=1}^3 \sqrt{p^2_i + m^2_i} + V_Y; ~~~V_Y= 
a \sum_{i=1}^3 | \vec{x_i} - \vec{x_T}| 
\end{equation}
where $m_i$ is the current mass, $a$  the string tension and
 $\vec{x_T}$ the Toricelli point. 
Our purpose is to obtain an approximate analytical form of the
eigenvalues of the Hamiltonian (\ref{EXACT}).  To a good approximation for 
the Y-junction \cite{BSNV}
  one can replace  $H$ by 
\begin{equation}
 H_0 = \sum_{i=1}^3 \sqrt{p^2_i + m^2_i} + 
\frac{a}{2}\left[ \sum_{i=1}^3| \vec{x_i} - \vec{R}| +
 \frac{1}{2} \sum_{i<j}^3 | \vec{x_i} - \vec{x_j}|\right]
\end{equation}
where $\vec{R}$ is the position of the center of mass.
The next step is to use the auxiliary field formalism
\cite{YUS} 
which allows to replace a semirelativistic by a nonrelativistic 
kinetic energy  
and  a linear by a quadratic confinement. The eigenvalue problem
becomes exactly solvable.  By minimizing with respect to the 
auxiliary fields one obtains a good approximation to the exact
mass. This is  \cite{BS}

\begin{equation}\label{mass3q}
M_0 = 6 \mu_0.
\end{equation}
where 
\begin{equation}\label{mu0}
\mu_0 = [ \frac{a}{3} Q (N+3)]^{1/2}, ~~ Q = 1/2 + \sqrt{3}/4.
\end{equation}
Here $N = 2 n + \ell$, as in a harmonic oscillator potential,
and represents the band number used in phenomenology.  
Adding perturbatively Coulomb-type and self-energy corrections to
the squared mass of Eq. (\ref{mass3q}) one obtains   
\begin{equation}\label{totalmass}
M^2_0 = 2 \pi \sigma (N+3) -  
\frac{4}{\sqrt{3}}\pi\sigma\alpha_s-\frac{12}{(2+\sqrt{3})}f\sigma.
\end{equation}
provided one makes the scaling $12 a Q = 2 \pi \sigma$
where $\sigma$ is the standard strength tension, $\alpha_s$ is the
strong coupling constant and $f$ a parameter varying between 3 and 4.

In the auxiliary field formalism, one expects that
$c_2\propto\mu^{-2}_0 $ and $c_4\propto\mu^{-2}_0$. 
Thus, using Eq. (\ref{mu0}), one
obtains
\begin{equation}\label{c2c4qm}
c_2=\frac{c^0_2}{N+3},\quad c_4=\frac{c^0_4}{N+3}.
\end{equation}
where the coefficients $c^0_2$ and $c^0_4$ have to be fitted.

As a matter of fact, the proof that 
the band number $N$ can be considered a good quantum number for baryons 
including both strange and nonstrange quarks is more involved
(for details see Ref. \cite{SBS}).

\begin{figure}[hb]
\begin{center}
\includegraphics[height=6 cm, width=8.5 cm]{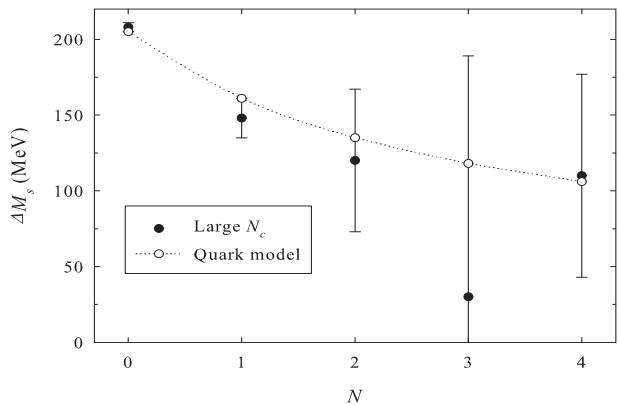}
\caption{$\Delta M_s$ from the $1/N_c$ expansion mass formula, 
Eq. (\ref{massoperator}) and the 
QM mass result (for details see Ref. \cite{SBS}).} \label{fig:strangemass}
\end{center}
\end{figure}

\subsection{Comparison of the two approaches}

In the real world we have $N_c = 3$. Thus we have to compare
the coefficient $c^2_1$ of Eq. (\ref{massoperator})  with 
$M^2_0/9$ where  $M^2_0$ comes from Eq. (\ref{totalmass}).
This comparison is made 
in Fig. \ref{fig:coef1} for the  
the  bands  N = 0,1,2,3,4 studied within the $1/N_c$ approach.
From the best fit one has 
$\sigma=0.163\pm0.004$~GeV$^2$, 
$\alpha_s=0.4$,  and $f=3.5$, as very standard values.
One can see a remarkable agreement between the two approaches.
In both cases 
the points follow the same straight line (Regge trajectory).
If the value of $\sigma$ is chosen in the common phenomenological interval
$\sigma=0.17\ (0.20)$~GeV$^2$ one obtains the shaded area.

For the upper part of Fig. \ref{fig:coef2and4} 
we chose $c^0_2=208\pm60$~MeV so that the large $N_c$ 
point at $N=1$, for which the uncertainty is minimal, is exactly
reproduced. This coefficient is related to the contribution of
the spin-orbit operator $O_2$ which turns out to be very small
in both approaches. For the lower part of Fig. \ref{fig:coef2and4}
a good fit to all points gave $c^0_4=1062\pm198$~MeV.
Note that  $c^0_4\gg c^0_2$ which indicates that the spin-spin contribution
dominates over the spin-orbit term. This justifies
the quark model assumption that the spin-spin is the dominant
contribution to the hyperfine interaction. Details of this comparison can
be found in Ref. \cite{SBMS}.

The comparison between the QM  and large $N_c$ results for $\Delta M_s$ 
is shown in Fig.  \ref{fig:strangemass}. 
The point at $N$ = 2 is from Ref. \cite{SBS}.
The points corresponding to $N$ = 0,1 and 3 are taken from Ref. \cite{GM}.
 Except for $N = 3$, the central values of $\Delta M_s$ in the large
$N_c$ approach are close to the quark model results. 
The QM results show a smooth behavior. This suggests that the  $N = 3$
point in the $1/N_c$ expansion must be re-analyzed. The accuracy of
the $1/N_c$ expansion results depends on the quality and quantity
of experimental data on strange baryons,
which is very scarce for various reasons \cite{PDG}. More data are highly
desired. 

\section{Conclusions}
The key tool in the comparison of QM and large $N_c$ approaches
is the band number $N$ which turns out to be a good and relevant
quantum number in the classification of baryons. It leads
to Regge trajectories where $M^2 \propto N$.   
The basic conclusion is that the large $N_c$ approach supports
the quark model assumptions as the relativistic kinetic energy, Y-junction
confinement, dominant spin-spin interaction, vanishing spin-orbit 
contribution, etc. At the same time, the QM can give some physical insight into
the coefficients $c_i$ and $d_i$ which encode the QCD dynamics. 
Similar studies are needed for heavy baryons.
\section*{Acknowledgments}

I am grateful to C. Semay, F. Buisseret and N. Matagne for an
enjoyable collaboration and to the MENU07 Conference organizers for their  
invitation. 








\end{document}